\documentclass[conference]{IEEEtran}
\IEEEoverridecommandlockouts
\usepackage{cite}
\usepackage{amsmath,amssymb,amsfonts}
\usepackage{algorithmic}
\usepackage{algorithm}
\usepackage{graphicx}
\usepackage{textcomp}
\usepackage{xcolor}
\usepackage{url}
\usepackage{booktabs}
\usepackage{multirow}
\usepackage{xspace}
\usepackage[hidelinks]{hyperref}
\usepackage{dirtree}
\usepackage{enumitem}
\usepackage{balance}
\usepackage{caption}
\usepackage{subcaption}

\usepackage{tikz}
\usetikzlibrary{positioning, arrows.meta}

\def\BibTeX{{\rm B\kern-.05em{\sc i\kern-.025em b}\kern-.08em
    T\kern-.1667em\lower.7ex\hbox{E}\kern-.125emX}}

\usepackage{xcolor}

\newcommand{\ie}{\emph{i.e.}, }
\newcommand{\eg}{\emph{e.g.}, }
\newcommand{\etc}{\emph{etc.\xspace}}

\begin{document}

\title{{MoQSplat}: Adaptive Progressive Streaming of \\3D Gaussian Splatting via MoQ}

\author{
\IEEEauthorblockN{
Emanuele Artioli$^1$\IEEEauthorrefmark{1},
Mohammadreza Ghafari$^1$\IEEEauthorrefmark{2},
Md Tariqul Islam$^1$\IEEEauthorrefmark{3},\\
Farzad Tashtarian\IEEEauthorrefmark{1},
Christian Rothenberg\IEEEauthorrefmark{3},
Christian Timmerer\IEEEauthorrefmark{1}
\thanks{The first three authors contributed equally to this work. The research was supported in part by the Austrian Federal Ministry for Digital and Economic Affairs, National Foundation for Research, Technology and Development, Christian Doppler Research Association with Project Christian Doppler Laboratory ATHENA (http://athena.itec.aau.at), French ANR PEPR NF-NAI project, No ANR-22-PEFT-0003, and partly by the French PIA project “Lorraine 
Université d’Excellence”, reference ANR-15-IDEX-04-LUE. Additional support was provided by Ericsson Telecomunicações Ltda., the São Paulo Research Foundation (FAPESP) under Grant No. 2021/00199-8 (CPE SMARTNESS), and in part by CAPES, Brazil (Finance Code 001).}
}

\IEEEauthorblockA{\textit{
\IEEEauthorrefmark{1}Christian Doppler Laboratory
ATHENA, Alpen-Adria Universität Klagenfurt, Austria}\\
\textit{\IEEEauthorrefmark{2}Université de Lorraine, CNRS, Inria, LORIA, F-54000 Nancy, France}\\
\textit{\IEEEauthorrefmark{3}Universidade Estadual de Campinas (UNICAMP), Brazil
}}
}

\maketitle

\begin{abstract}
3D Gaussian Splatting (3DGS) enables photorealistic novel view synthesis, but transmitting gigabyte-scale scene data remains challenging for immersive applications. Traditional HTTP Adaptive Streaming over TCP introduces Head-of-Line (HOL) blocking and coarse segmenting ill-suited to fine-grained 3DGS delivery. We propose \texttt{MoQSplat}, which maps 3DGS content onto the Media over QUIC (MoQ) transport hierarchy. \texttt{MoQSplat} partitions scenes into spatial Tracks, clusters splats into semantically coherent Groups, and constructs progressive-quality Subgroups mapped to independent QUIC streams to eliminate connection-level HOL blocking. Using a stateless, subscriber-driven adaptation loop, clients dynamically request spatial regions and quality tiers based on six degrees of freedom (6-DoF) frustum visibility, distance, and foveal alignment. We evaluate the core components on a prototype implementation, showing that opacity-based pruning outperforms scale-based pruning for progressive delivery.
The source code is available at \url{https://github.com/emanuele-artioli/MoQSplat}.
\end{abstract}


\noindent{\footnotesize \textcopyright~2026 IEEE. Personal use of this material is permitted. Permission from IEEE must be obtained for all other uses, in any current or future media, including reprinting/republishing this material for advertising or promotional purposes, creating new collective works, for resale or redistribution to servers or lists, or reuse of any copyrighted component of this work in other works.}
\vspace{1mm}

\section{Introduction}

Photorealistic 3D scene delivery is critical for immersive applications like VR telepresence and digital twins. 3D Gaussian Splatting (3DGS)~\cite{kerbl20233d} achieves real-time, high-fidelity rendering by using explicit collections of Gaussian primitives instead of computationally intensive neural inference like NeRFs~\cite{Mildenhall2021nerf}. However, capturing complex view-dependent specularities requires up to 3rd-degree Spherical Harmonics (45 parameters per primitive)~\cite{hasssan2026dcsharp}. Consequently, realistic 3DGS scenes comprise millions of splats exceeding several gigabytes, imposing severe bandwidth and transmission challenges for real-time delivery over standard networks.

To deliver 3DGS efficiently, recent frameworks adapt concepts from HTTP Adaptive Streaming (HAS) formats (\eg MPEG-DASH) and Scalable Video Coding (SVC)~\cite{dashsvc2012}. For example, LapisGS~\cite{shi2025lapisgs} employs opacity optimization for progressive layering, LTS~\cite{sun2025lts} combines spatial tiles with DASH segments, and L3GS~\cite{tsai2025l3gs} uses viewport prediction for object scheduling. However, operating over TCP introduces Head-of-Line (HOL) blocking, where packet loss stalls all subsequent data retransmission~\cite{ravuri2023adaptive}. This combination of TCP's rigid reliability and HAS's coarse segment granularity is ill-suited to fine-grained 3DGS data.

\begin{figure}
    \centering
       \vspace{-4mm}
    \includegraphics[width=1\linewidth]{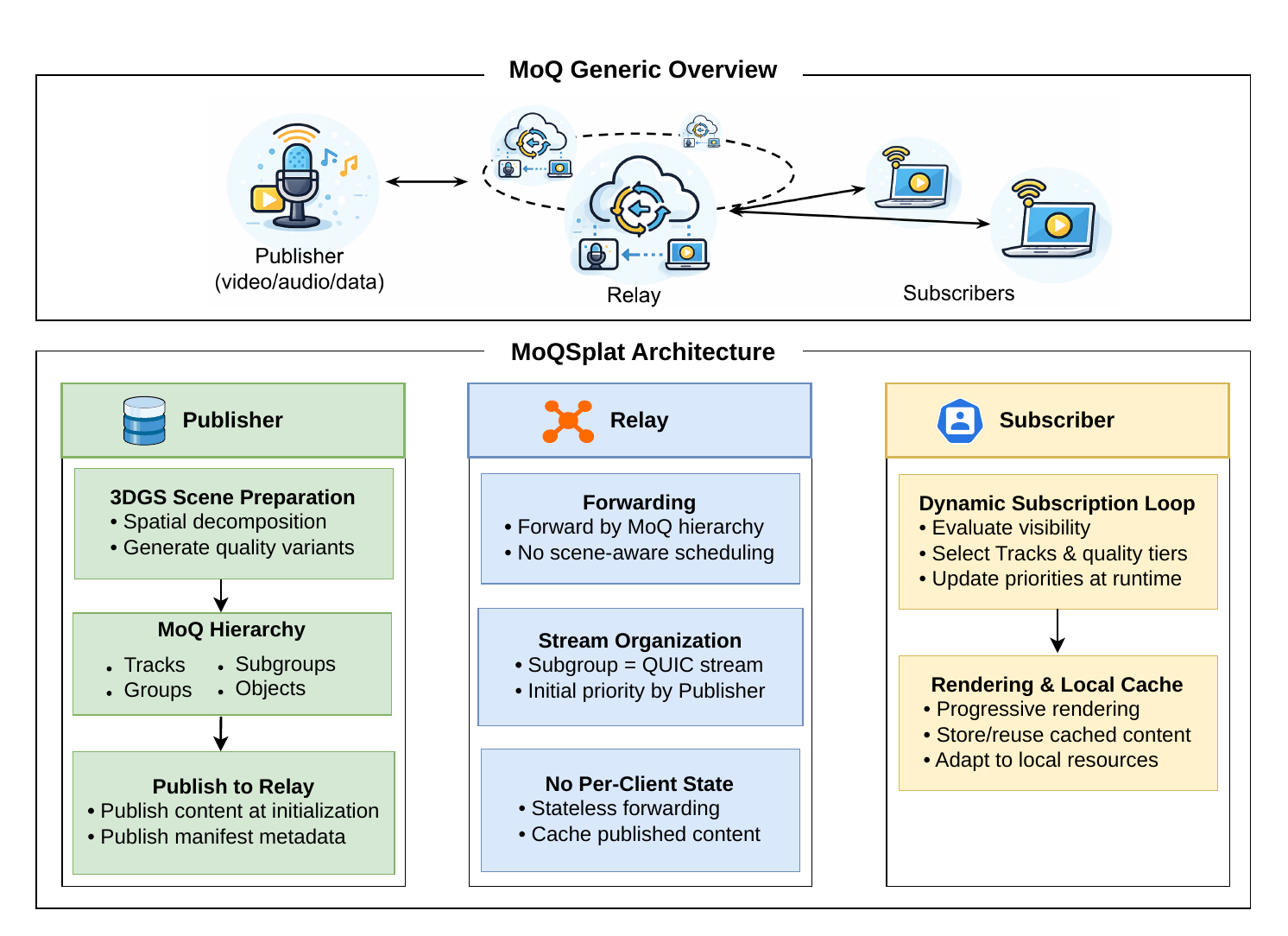}
   \vspace{-4mm}
    \caption{{Overview of the MoQ and \texttt{MoQSplat} architecture.}}
    \label{fig:pub-sub}
    \vspace{-4mm}
\end{figure}

While HTTP/3 over QUIC mitigates HOL blocking via multiplexed UDP streams \cite{sidhu2025video}, standard HTTP/3 client-pull segmenting still inflates latency. To address this, Media over QUIC (MoQ)~\cite{moq-draft} is emerging as an IETF standard for low-latency interactive media. MoQ utilizes a Publish-Subscribe (Pub/Sub) architecture with hierarchical Tracks, Groups, Subgroups, and Objects deliverable via intermediate Relays (Fig.~\ref{fig:pub-sub}). Although MoQ streaming has been explored for point clouds~\cite{freeman2026implicit,nemeth2024transmitting}, it has not been applied to 3DGS's anisotropic scales, opacities, and spherical harmonics. MoQ's structure maps naturally to 3DGS, enabling independent streaming and priority-aware delivery of spatial regions and quality tiers without connection-level HOL blocking.

In this paper, we propose \texttt{MoQSplat}, an adaptive progressive streaming architecture designed to address the challenges of transmitting 3D Gaussian splats. \texttt{MoQSplat} maps Gaussian splatting onto MoQ's hierarchical content structure to enable adaptive progressive streaming through subscriber-driven quality selection. The main contributions are as follows:
\begin{enumerate}[leftmargin=*,nosep]
    \item MoQ hierarchy mapping for 3DGS: We map 3DGS onto MoQ's four-level content hierarchy, detecting semantic objects and defining quality refinements into distinct Subgroups.
    \item Subscriber-driven adaptation: We introduce a fully stateless client-side subscription model. The client independently calculates utility based on frustum alignment and Euclidean distance, dynamically adjusting layer granularity via \texttt{SUBSCRIBE} commands.
    \item Framework implementation: We outline the end-to-end Python-based \texttt{MoQSplat} implementation built over the customized \texttt{moq-lite}\footnote{\url{https://github.com/moq-dev/moq}, last accessed: 16 September 2026.} Pub/Sub framework and PyTorch-based \texttt{gsplat}\footnote{\url{https://github.com/nerfstudio-project/gsplat}, last accessed: 16 September 2026.} renderer.
\end{enumerate}

\begin{figure*}
    \centering
    \includegraphics[width=1\linewidth]{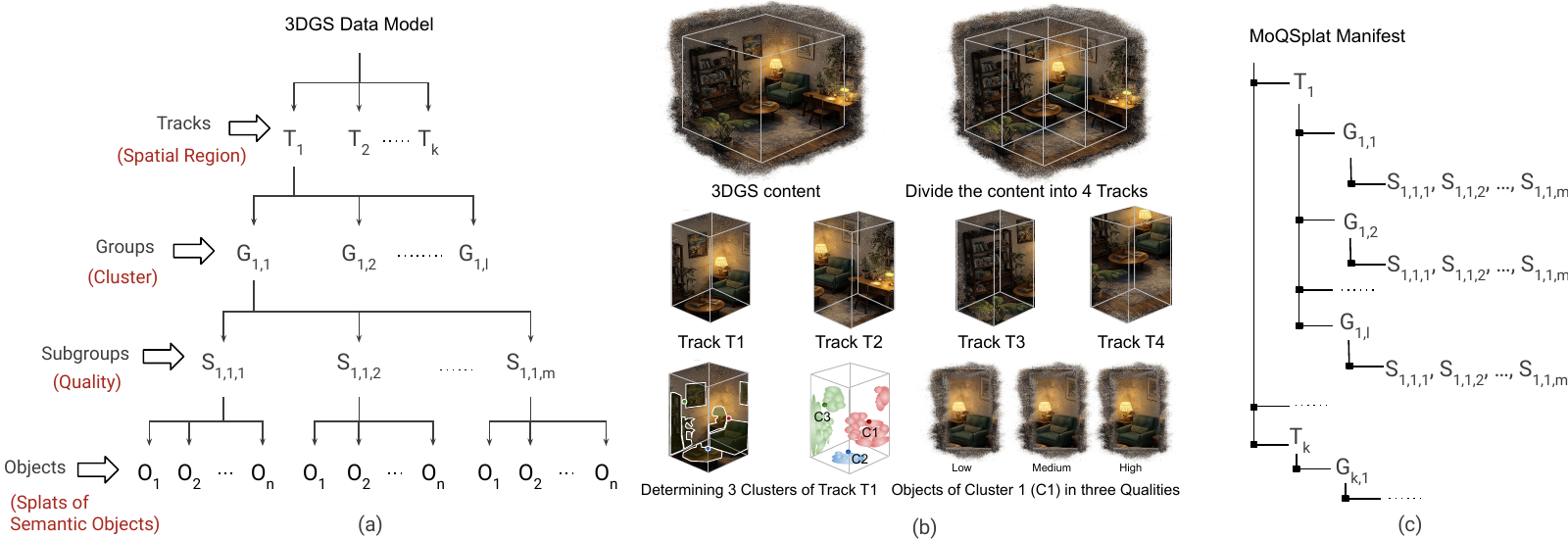}
    \vspace{-4mm}
    \caption{{MoQSplat hierarchical data model: (a) four-level MoQ mapping of a 3DGS scene; (b) example of spatial partitioning into Tracks, semantic clustering, and multi-quality rendering of an indoor scene; (c) proposed MoQSplat Catalog (Manifest).}}
    \label{fig:arch}
    \vspace{-4mm}
\end{figure*}

\section{MoQSplat Design}
\label{sec:design}
\subsection{MoQ hierarchy mapping for 3DGS}
\texttt{MoQSplat} comprises three primary components: \textit{(1)} Publisher, \textit{(2)} MoQ-compliant content-agnostic Relay (optional), and \textit{(3)} Subscriber, interconnected within a MoQ architecture, as shown in Fig.~\ref{fig:pub-sub}. 
A core principle of this design is that quality adaptation is Subscriber-driven, i.e., the Publisher exposes all spatial and quality variations, while the Relay serves as a content-agnostic forwarding and caching entity, and the Subscriber dynamically requests and prioritizes streams to match the viewer's current viewport. 
This aligns with on-demand static scene delivery and supports efficient multi-subscriber content distribution. To support progressive delivery and selective subscription, \texttt{MoQSplat} maps the 3DGS scene content onto a structured four-level MoQ hierarchy, as illustrated in Fig.~\ref{fig:arch}~(a):

\textit{Track.} At the coarsest granularity, the global 3DGS content is partitioned into multiple non-overlapping spatial regions, referred to as \emph{Tracks} (Fig.~\ref{fig:arch}~(b)). 
This decomposition improves scalability for large-scale environments containing millions of Gaussian splats by reducing memory consumption and computational complexity. 
Clients subscribe only to Tracks corresponding to the regions currently visible in their viewport, thereby avoiding the transmission of all splats. 
Existing spatial subdivision techniques, such as \texttt{block partitioning} strategies commonly adopted in point cloud~\cite{Ghafari2025PC, Qin2025PC}, can be employed to construct these Tracks efficiently.

\textit {Group.}
Within each Track, splats are grouped into semantically coherent object clusters. Grouping can be performed directly on splat attributes using spatial clustering (\eg \texttt{K-Means}~\cite{Gao2025ProtoGS}, \texttt{DBSCAN}~\cite{Chen2024FreeGaussian}) or via 2D/3D object detection mapped back to the 3D domain. Semantic awareness is vital: naive spatial partitioning risks splitting single objects across clusters, causing visual artifacts during independent streaming. Grouping ensures coherent object-level associations, where spatially adjacent objects may form unified clusters (\eg \(C1\)--\(C3\) in Fig.~\ref{fig:arch}~(b)).

\textit {Subgroup.}
Each Group is partitioned into Subgroups representing quality levels ($S_1, \dots, S_m$) using two efficient heuristics:
\textit{1) Independent Geometric Heuristics} - partition the splat population by ranking spatial saliency (volume and isolation) to ensure base layer $S_1$ minimizes visual holes. Subgroup boundaries are configured via percentage thresholds or fixed cardinality budgets, and can integrate pruning~\cite{Papantonakis20243DGS} or compression~\cite{Bagdasarian2025Compression}.
\textit{2) Dependent Attribute-Based Partitioning} - splits splat attributes across layers. Base layer $S_1$ transmits geometry and 0-th order Spherical Harmonics (SH) diffuse color, while higher-order SH coefficients are deferred to enhancement layers. This introduces a decoding dependency requiring $S_1$ geometry before enhancement layers can be rendered.

\textit {Object.}
At the finest granularity, each Subgroup consists of multiple Objects that represent the transmission units carried over the network. These Objects encapsulate the individual Gaussian splat attributes, including geometry, opacity, Spherical Harmonics coefficients, \etc, which naturally represent semantic objects of the associated cluster (\eg a couch in cluster \(C1\), illustrated in Fig.~\ref{fig:arch}~(b)). 

\subsection{Publish-Subscribe Design}
\textbf{Publisher}. 
The Publisher is responsible for the content preprocessing, hierarchy mapping, and session initialization. 
It partitions the 3DGS scene content spatially and semantically to fit the defined four-level MoQ hierarchy, then generates a dedicated manifest containing scene metadata. 
This manifest details the Track identifiers along with their spatial bounding volumes, as well as the Group and Subgroup organizational structure (as shown in Fig.~\ref{fig:arch}~(c)).
Each Subgroup uses an independent QUIC stream, enabling progressive delivery from base to enhancement quality layers while avoiding connection-level HOL blocking. 
During initialization, the Publisher makes the manifest and all precomputed spatial and quality variants available to the Relay (if present), where they may be cached for the duration of the session.

Since quality adaptation is subscriber-driven, the Publisher does not actively manage runtime stream priorities. 
Instead, it embeds structural metadata in the manifest (such as Group 3D centroids and Subgroup progressive quality levels) to facilitate subscriber-viewport adaptation calculations. 
To enable progressive delivery, the Publisher assigns static baseline priorities to Subgroups.
Specifically, these priorities are set to low values (representing high priority, as MoQ priorities are integers in the interval $[0, 255]$ where $0$ is the most prioritized) for the base layer, and progressively higher values (lower priority) for each subsequent enhancement layer. 
Consistent with MoQ's prioritization model, the scheduling engine evaluates subscriber-assigned subscription priorities first, using publisher-assigned priorities only as a tie-breaker.

\textbf{Subscriber}. 
Unlike HTTP-based adaptive streaming, \texttt{MoQSplat} Subscriber performs adaptation through stream prioritization rather than segment selection. It exploits application-layer scheduling over QUIC's multiplexed streams to allocate transmission capacity to the most relevant streams based on the Subscriber priorities.
Upon session initialization, the Subscriber retrieves the publisher-generated manifest.
Using the spatial metadata embedded within it, the Subscriber filters active subscriptions at the coarsest granularity, subscribing only to the Tracks that intersect the user's viewport.
Within these active Tracks, the Subscriber periodically computes dynamic subscription priorities for the underlying Subgroups. 
The Subscriber can employ any prioritization algorithm to determine these values. 
For instance, it can utilize viewport-aware heuristics that ingest real-time 6-DoF telemetry alongside structural metadata (such as the Group 3D centroids and Subgroup quality levels defined earlier in this section) to prioritize nearby and foveated objects.
These priority updates are transmitted upstream via standard MoQ subscribe update messages. 
This 
enables accelerating streams for visually important objects while throttling less relevant ones without requiring connection 
re-establishment. 

This Subscriber-driven prioritization scheme serves as the 3D-viewport-aware counterpart to traditional adaptive bitrate (ABR) algorithms in 2D video streaming. 
Accordingly, designing and evaluating advanced viewport-aware scheduling heuristics under highly dynamic network and telemetry conditions represents a promising direction for future research. If the Subscriber does not employ any prioritization algorithm, every Subgroup will tie for the same Subscriber priority, falling back entirely to these progressive publisher-defined priorities. Consequently, all base layers across all visible clusters are sent simultaneously, followed by all first enhancement layers, and so forth, establishing a baseline delivery mode.

To cope with local resource constraints, the Subscriber can employ a hierarchical memory management scheme across VRAM (where rendered tensors reside), system memory, and storage. Layers associated with low-priority or occluded objects are demoted to lower memory, while previously cached data can be promoted back to VRAM for rendering when object importance increases. This reduces redundant network transfers while maintaining rendering performance and efficient resource utilization.

\section{MoQSplat Implementation}
To demonstrate the feasibility and evaluate the performance of the \texttt{MoQSplat} architecture, we developed a functional prototype. Rather than comprehensively realizing all the architectural branches described in Section~\ref{sec:design}, the prototype implements a specific, concrete configuration of the pipeline to showcase the viability of the overarching framework. 

The implementation's content decomposition pipeline processes the 3DGS content to structure it into a single-track MoQ hierarchy. The system consists of three runtime components: the Python-based Publisher that serializes and transmits the hierarchy, a standard, content-agnostic Relay, and the Subscriber that drives viewport-steered adaptation during streaming.

\textbf{Publisher Implementation}.
In our initial implementation, we consider only a single Track that covers the entire 3DGS content. To organize the content, the following steps must be applied on the Publisher prior to streaming:

\textit{1) Semantic Object Detection:} To extract Groups, the server renders the content from \(N\) different viewpoints and performs 2D object detection on the generated images. The rendered images are processed using \texttt{YOLOv26x}~\cite{Cheng2024yolo} to identify objects in each view. Since the content is observed from \(N\) viewpoints, multiple bounding boxes may correspond to the same underlying semantic object. To associate these detections with Gaussian splat regions, the server estimates a confidence region over the corresponding splats to robustly localize the detected objects. To improve robustness, each YOLO-detected bounding box is optionally expanded by a predefined padding factor, ensuring conservative spatial coverage of the underlying detected object.

\textit{2) Merge Voting:} A semantic object may not be fully covered by a single viewport, thus, multiple viewports were used. However, when an object is observed from multiple viewpoints, redundant detections are merged to form a single bounding box associated with the object, ensuring complete coverage of the splats representing it. This is achieved either by computing a 3D Jaccard index (IoU) over the corresponding Gaussian splats or by measuring the centroid proximity of the associated splats. In this way, repeated observations of the same semantic object across different views (\eg a bicycle detected from multiple angles) are combined into a single unified Gaussian representation.

\textit{3) Clustering:} The resulting multi-view boundary box detections are then projected back onto the splat space, enabling the construction of semantically meaningful Groups. Specifically, Gaussian splats are clustered according to object centroids and spatial proximity, such that nearby splats are assigned to the same Group, yielding coherent object-level clusters. To perform this grouping, \texttt{DBSCAN}~\cite{Chen2024FreeGaussian} is applied over the detected object centroids, treating objects as spatial points, and forming multiple clusters. Some clusters may include multiple semantic objects if they are spatially nearby.

\textit{4) Progressive Subgroup Construction:} To implement progressive quality levels, the prototype realizes the \emph{Independent Geometric Heuristics} outlined in our design. Specifically, the Publisher partitions the splat population by applying geometry pruning based on scale and opacity proxy measures, using a step size of 20\% to construct five progressive quality Subgroups.
This pipeline enables progressive and adaptive 3DGS streaming through hierarchical Group and Subgroup delivery driven by subscriber-driven viewport adaptation.

\textit{5) Pre-Processing Complexity:} Performing multi-view rendering across $N$ viewpoints and 2D object detection via \texttt{YOLOv26x} has a computational complexity of $\mathcal{O}(N \cdot T_{\text{YOLO}})$. Projecting detected 2D bounding boxes back into 3D Gaussian splat space and applying \texttt{DBSCAN} clustering scales as $\mathcal{O}(M \log M)$ over $M$ Gaussian centroids, while opacity sorting for progressive subgroup creation requires $\mathcal{O}(M \log M)$. Crucially, this pre-processing pipeline operates entirely offline as a one-time content ingestion phase for static 3DGS-on-demand scenes, eliminating real-time pre-processing latency during streaming.

\textit{MoQ Hierarchy Extension.}  
The Publisher is implemented in Python and interfaces with \texttt{moq-lite}, a Rust-based Pub/Sub framework over QUIC and WebTransport, through a shared Rust library exposed via ctypes bindings. The Publisher is responsible for serializing the precomputed 3DGS hierarchy and publishing the resulting content to the Relay during session initialization. 
Although the current MoQ transport specification includes Subgroups~\cite{moq-draft}, the \texttt{moq-lite} version used here supported only a three-level hierarchy (Track, Group, Object); we therefore extended it with an explicit Subgroup layer for progressive Gaussian delivery, resulting in Track, Group, Subgroup, Object.
Each Subgroup corresponds to an independent QUIC stream and is assigned an initial Publisher priority according to its quality level. Gaussian attributes are encoded as raw float32 arrays, since a standardized 3DGS compression codec has not yet been finalized.
At initialization, the Publisher serializes the manifest (\ie Catalog) as a JSON-based MoQ Track containing the precomputed hierarchy.

\textbf{Relay Implementation}. 
\texttt{MoQSplat} adopts \texttt{moq-relay}, the Rust-based reference implementation of \texttt{moq-lite}, as a content-agnostic, priority-aware QUIC forwarding service. The Relay forwards content structured into Tracks, Groups, Subgroups, and Objects, as defined by our Pub/Sub shared library, and relays upstream control messages (\eg \texttt{SUBSCRIBE\_UPDATE}) for dynamic priority adjustment. As  MoQ transport continues to evolve toward RFC standardization~\cite{moq-draft}, future revisions may be needed to align with the finalized specification.



\textbf{Subscriber Implementation}. The Subscriber is implemented in Python, utilizing PyTorch with the \texttt{gsplat} library to execute hardware-accelerated rasterization, and interfacing with \texttt{moq-lite} through the shared Rust library via ctypes bindings.
The Subscriber instantiates the client-driven quality adaptation through the following components:

\textit{1) Subscription Hierarchy and State Machine:} Upon establishing a connection, the Subscriber retrieves the Publisher’s MoQ Catalog and reconstructs the available Track, Group, and Subgroup hierarchy. To manage subscription lifecycles, the Subscriber maintains a lightweight, application-level state machine for each Subgroup (\eg \texttt{MISSING}, \texttt{DOWNLOADING}, or \texttt{CACHED}) to track local subscription and caching status.
If a transport loss, drop, or timeout occurs while a Subgroup is in the \texttt{DOWNLOADING} state, the subscriber re-evaluates the stream after a timeout threshold, transitioning its state back to \texttt{MISSING} to trigger stream re-subscription via a new \texttt{SUBSCRIBE} message. Meanwhile, the renderer gracefully falls back to displaying the currently available base layer ($S_1$) and lower-index enhancement layers already cached in VRAM, avoiding visual stalls or freezing.

\textit{2) Viewport-Steered Prioritization Loop:} The core transport interactions are governed by an asynchronous control loop (\eg 100\,ms intervals) to dynamically compute and update stream priorities.
During each iteration, the Subscriber extracts the current 6-DoF pose and calculates a priority score, denoted as $\Delta P$, for all cataloged Groups:
\begin{equation}
    \small
    \Delta P = W_d \cdot \max \left(0, 1 - \frac{\min(D, D_{\text{max}})}{D_{\text{max}}} \right) + W_a \cdot \max(0, \cos \theta)
\end{equation}
where $D$ is the Euclidean distance from the subscriber's camera position to the Group's 3D centroid, $\theta$ is the angle between the camera's forward viewing vector and the vector pointing from the camera to the Group's centroid, $D_{\text{max}}$ is the expected maximum viewing distance, and $W_d$ and $W_a$ represent scaling weights prioritizing proximity and centralized foveal alignment, respectively. 
Upon computing $\Delta P$, the subscriber evaluates its local state machine. If a required Subgroup's state is \texttt{MISSING}, a new \texttt{SUBSCRIBE} message is issued to the Relay with the calculated $\Delta P$. Crucially, if the Subgroup is already \texttt{DOWNLOADING}, the subscriber dispatches a \texttt{SUBSCRIBE\_UPDATE} to alter the stream's priority on the transport wire without closing the connection. 
Incoming Objects are unpacked and appended to a local memory storage pool; once a Subgroup's Objects are fully received, its state transitions to \texttt{CACHED}.

\textit{3) Watermark-Based Memory Swapping:} To implement the hierarchical memory management scheme proposed in the design, the prototype Subscriber features a two-tier, watermark-based memory swap strategy between high-performance GPU VRAM and system RAM. The Subscriber establishes a fixed GPU memory budget, $M_{\text{max}}$, and a high-watermark threshold (\eg $0.9 \times M_{\text{max}}$). If the active splat memory footprint exceeds this threshold, the Subscriber protects all base-layer Subgroups from eviction and sorts enhancement layers by their current $\Delta P$ priority modifier. The lowest-priority enhancement layers are evicted from GPU VRAM, demoting their tensors to the \texttt{CACHED} state in system RAM. When a user changes viewports and a cached object's $\Delta P$ rises, the Subscriber promotes the corresponding data back to VRAM, avoiding redundant network re-transmissions and rendering stalls.

\begin{figure*}[t]
    \centering
    \includegraphics[width=1\linewidth]{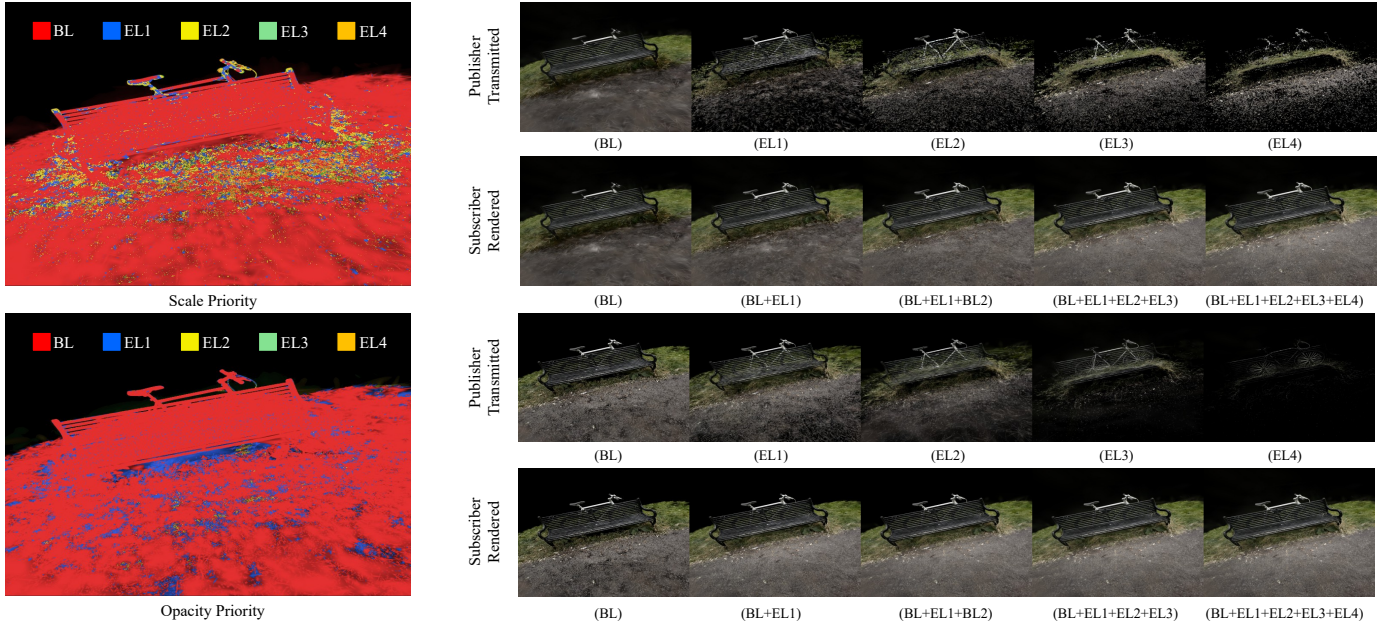}
    \caption{{\textit{Bench--Bicycle} cluster under progressive quality enhancement achieved via opacity- and scale-based pruning. The Publisher transmits enhancement layers, each containing 20\% of the total splats. The Subscriber cumulatively aggregates these layers to progressively improve visual quality. Note: BL = Base Layer; EL1–EL4 = Enhancement Layers 1–4.}}
    \label{fig:yolo_splats}
 \end{figure*}

\begin{figure}[t]
    \centering
    \includegraphics[width=1\linewidth]{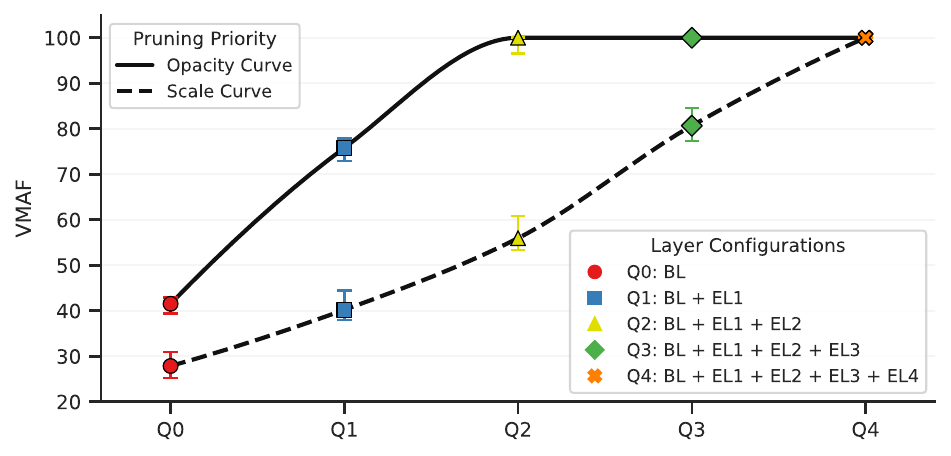}
    \caption{{Progressive quality enhancement measured by VMAF.}}
    \label{fig:vmaf}
    \vspace{-4mm}
\end{figure}

\section{Preliminary Evaluation Results}
To validate the proposed \texttt{MoQSplat} pipeline, we evaluate it in two stages: first, we assess Publisher-side content preparation (offline) following the implementation described earlier; then, we evaluate end-to-end streaming delivery under varying network conditions. The \textit{Bicycle} content from the Mip-NeRF360 dataset~\cite{barron2022mipnerf360} is used as a common reference across both stages.


\textbf{Publisher-Side Content Preparation}.
VMAF was used as the quality metric, computed by smoothly orbiting the target 3DGS cluster (\textit{Bench–Bicycle}) over a $\pm90^\circ$ azimuth range while maintaining a fixed distance, elevation, and camera intrinsics. Each sequence lasted 10s at 30 fps (300 frames), with the unpruned model serving as the reference for evaluating the pruned models.
Publisher-side experiments were conducted on an Ubuntu 22.04 cluster with Intel Xeon Gold 5218 CPUs @ 2.30 GHz and dual NVIDIA Quadro GV100 GPUs. From Fig.~\ref{fig:yolo_splats} and Fig.~\ref{fig:vmaf}, the following observations can be drawn:

\textit{Group.} The \textit{Bench--Bicycle} objects are clustered from the full \textit{Bicycle} scene (illustrated in Fig.~\ref{fig:yolo_splats}), showing the effectiveness of the proposed Group strategy in identifying semantically meaningful clusters. This enables adaptive progressive streaming by transmitting only relevant splats to the Subscriber.

\textit{Subgroup.} Both opacity- and scale-based pruning improve visual quality as additional splats are progressively received by the Subscriber, as reflected in Fig.~\ref{fig:yolo_splats}. However, opacity-based pruning consistently outperforms scale-based pruning by achieving higher perceptual quality with the same amount of rendered splats, as shown in Fig.~\ref{fig:yolo_splats} and Fig.~\ref{fig:vmaf}. These results demonstrate the effectiveness of opacity-based prioritization for progressive rendering, enabling Subscriber to achieve satisfactory visual quality in the requested viewport without waiting to receive the full set of splats.


\textit{Enhancement Layer Analysis:} Although each base/enhancement layer (BL, EL1-EL4) transmitted by the Publisher contains an equal amount of splats (20\% of the total splats in each base/enhancement layer), their perceptual contributions differ significantly, as shown in Fig.~\ref{fig:yolo_splats}. This highlights the importance of which splats are transmitted first. The cumulative combination of layers progressively improves reconstruction quality from Q0 to Q4 (Fig.~\ref{fig:vmaf}), where the BL provides the initial rendering and subsequent ELs refine the quality. However, gains diminish under opacity-based pruning as quality saturates beyond Q2 (Fig.~\ref{fig:vmaf}), with additional splats yielding marginal improvements. Although these additional splats appear less useful at typical viewing distances, their contribution may become more noticeable at closer viewpoints near the \textit{bench}.

\textbf{End-to-End Streaming Delivery}.
To validate the progressive delivery behavior of \texttt{MoQSplat}, we present initial results from a streaming session using the \textit{Bicycle} 3DGS content under different bandwidth constraints. The Publisher preprocesses the content into the MoQ hierarchy and publishes it to the Relay during initialization. Congestion is handled implicitly by the Relay, and higher-priority Subgroups are delivered first, enabling progressive scene refinement under limited bandwidth.
\begin{table}[t]
\centering
\caption{\small{Subscriber-driven Delivery Performance ---\textit{Bicycle} Scene}}
\label{tab:streaming_results}
\begin{tabular}{cccc}
\toprule
\textbf{Link BW} & \textbf{Subgroups} & \textbf{Data Received} & \textbf{Splats} \\
\textbf{(Mbps)} & \textbf{Delivered} & \textbf{(MB)} & \textbf{Rendered (Final)} \\
\midrule
10  & 397   & 37.4  & 224,710   \\
30  & 2,432 & 337.8 & 1,400,909 \\
50  & 2,628 & 363.2 & 1,491,950 \\
100 & 2,853 & 390.6 & 1,512,214 \\
\bottomrule
\end{tabular}
\vspace{-6mm}
\end{table}
Table~\ref{tab:streaming_results} shows subscriber-driven delivery metrics across four bandwidth conditions. At 10\,Mbps, the Subscriber receives 397 Subgroups (37.4\,MB), corresponding to approximately 225K Gaussian splats and yielding a coarse but continuously renderable scene representation. Increasing the available bandwidth to 30\,Mbps substantially improves delivery performance, with the number of received Subgroups increasing from 397 to 2,432 and the reconstructed splats growing from approximately 225K to 1.4M. Further increasing the bandwidth to 50\,Mbps results in a more modest improvement, reaching 2,628 delivered Subgroups and approximately 1.49M reconstructed splats. Finally, at 100\,Mbps, the Subscriber receives 2,853 Subgroups and approximately 1.51M splats, representing only a small increase over the 50\,Mbps case despite a twofold increase in available bandwidth.
Overall, these preliminary results demonstrate that \texttt{MoQSplat} maintains progressive scene refinement across diverse network conditions through priority-driven Subgroup delivery.

\section{Conclusions}
In this paper, we introduced \texttt{MoQSplat}, the first framework to map 3D Gaussian Splatting (3DGS) content onto the Media over QUIC (MoQ) hierarchical content structure. By partitioning scenes into spatial Tracks, semantically coherent Groups, and progressive-quality Subgroups mapped to MoQ Objects, \texttt{MoQSplat} enables prioritized delivery of spatial regions and quality layers, with Subscriber-driven adaptation avoiding per-subscriber state at the Publisher or Relay. Preliminary results show opacity-based pruning achieves strong visual quality with minimal bandwidth, and priority-driven delivery sustains progressive refinement across network conditions.
This work represents a first step toward MoQ-based 3DGS streaming. While the current prototype and evaluation are preliminary, several limitations remain across content preparation and delivery. \texttt{MoQSplat} demonstrates the feasibility of mapping 3DGS onto the MoQ hierarchy, establishing a foundation for more comprehensive evaluation, refinement, and optimization in future work.

\balance
\bibliographystyle{IEEEtran}
\bibliography{references}
\end{document}